\documentclass[fleqn,usenatbib]{mnras}

\usepackage{newtxtext,newtxmath}

\usepackage[T1]{fontenc}

\DeclareRobustCommand{\VAN}[3]{#2}
\let\VANthebibliography\thebibliography
\def\thebibliography{\DeclareRobustCommand{\VAN}[3]{##3}\VANthebibliography}

\usepackage{graphicx}	% Including figure files
\usepackage{amsmath}	% Advanced maths commands

\usepackage{multirow}
\usepackage{comment}
\usepackage{lipsum}
\usepackage{rotating}

\usepackage{scalerel}
\usepackage{tikz}
\usetikzlibrary{svg.path}
\definecolor{orcidlogocol}{HTML}{A6CE39}
\tikzset{
  orcidlogo/.pic={
    \fill[orcidlogocol] svg{M256,128c0,70.7-57.3,128-128,128C57.3,256,0,198.7,0,128C0,57.3,57.3,0,128,0C198.7,0,256,57.3,256,128z};
    \fill[white] svg{M86.3,186.2H70.9V79.1h15.4v48.4V186.2z}
                 svg{M108.9,79.1h41.6c39.6,0,57,28.3,57,53.6c0,27.5-21.5,53.6-56.8,53.6h-41.8V79.1z M124.3,172.4h24.5c34.9,0,42.9-26.5,42.9-39.7c0-21.5-13.7-39.7-43.7-39.7h-23.7V172.4z}
                 svg{M88.7,56.8c0,5.5-4.5,10.1-10.1,10.1c-5.6,0-10.1-4.6-10.1-10.1c0-5.6,4.5-10.1,10.1-10.1C84.2,46.7,88.7,51.3,88.7,56.8z};
  }
}
\newcommand\orcidicon[1]{\href{https://orcid.org/#1}{\mbox{\scalerel*{
\begin{tikzpicture}[yscale=-1,transform shape]
\pic{orcidlogo};
\end{tikzpicture}
}{|}}}}

\title[Earth-size formation in Kepler-1647]{Formation of Earth-sized planets within the Kepler-1647 System Habitable Zone}

\author[G. O. Barbosa et al.]{
G. O. Barbosa,$^{1,2}$\thanks{E-mail: gerson.barbosa@inpe.br (GOB)}\orcidicon{0000-0002-1147-2519}\,
O. C. Winter,$^{2}$\thanks{E-mail: othon.winter@unesp.br (OCW)}\orcidicon{0000-0002-4901-3289}\,
A. Amarante,$^{2,3,4}$ \thanks{E-mail: andre.amarante@ifsp.edu.br (AA)}\orcidicon{0000-0002-9448-141X}\,
E. E. N. Macau$^{1, 5}$ \thanks{E-mail: elbert.macau@unifesp.br (EENM)}\orcidicon{0000-0002-6337-8081}\
\\
$^{1}$National Institute for Space Research (INPE), Laboratório de Computação Aplicada,
             São José dos Campos, SP 12227-010,  Brazil.\\
$^{2}$São Paulo State University (UNESP), Grupo de Dinâmica Orbital e Planetologia, 
             Guaratinguetá, SP 12516-410, Brazil.\\
$^{3}$State University of Mato Grosso do Sul (UEMS), Cassil\^andia, MS 79540-000, Brazil.\\
$^{4}$Federal Institute of Education, Science and Technology of S\~ao Paulo (IFSP), Cubat\~ao, SP 11533-160, Brazil.\\
$^{5}$Federal University of São Paulo (UNIFESP), Institute for Science and Technology, 
             São José dos Campos, SP 12247-014, Brazil.
}
\date{Accepted 2021 April 20. Received 2021 April 20; in original form 2021 March 10}
\pubyear{2021}

\begin{document}
\label{firstpage}
\pagerange{\pageref{firstpage}--\pageref{lastpage}}
\maketitle

\begin{abstract}
The \textit{Kepler-1647} is a binary system with two Sun-type stars ($\approx$1.22 and $\approx$0.97 $M_\odot$). It has the most massive circumbinary planet ($\approx$1.52 $M_{Jup}$) with the longest orbital period ($\approx$ 1,107.6 days) detected by the \textit{Kepler} probe and is located within the habitable zone (HZ) of the system. In this work, we investigated the ability to form and house an Earth-sized planet within its HZ. First, we computed the limits of its HZ and performed numerical stability tests within that region. We found that HZ has three sub-regions that show stability, one internal, one co-orbital, and external to the host planet \textit{Kepler-1647b}. Within the limits of these three regions, we performed numerical simulations of planetary formation. In the regions inner and outer to the planet, we used two different density profiles to explore different conditions of formation. In the co-orbital region, we used eight different values of total disk mass. We showed that many resonances are located within  regions causing much of the disc material to be ejected before a planet is formed. Thus, the system might have two asteroid belts with \textit{Kirkwood gaps}, similar to the Solar System’s main belt of asteroids. The co-orbital region proved to be extremely sensitive, not allowing the planet formation, but showing that this binary system has the capacity to have Trojan bodies. Finally, we looked for regions of stability for an Earth-sized moon. We found that there is stability for a moon with this mass up to 0.4 Hill’s radius from the host planet.
\end{abstract}


%%%%%%%%%%%%%%%%%%%%%%%%%%%%%%%%%%%%%%%%%%%%%%%%%%

%%%%%%%%%%%%%%%%% BODY OF PAPER %%%%%%%%%%%%%%%%%%

\section{Introduction}
\label{sec:intro}

Often several exoplanets are discovered and confirmed in the most diverse dynamic conditions. Planets in binary systems are one of those exotic cases. Today they count 150 exoplanets in 102 systems (current data from the catalog presented in \cite{schwarz2016new}). These exoplanets are distributed in two types of orbits; (1) \textit{P-type}: systems where the planet is orbiting the binary pair. It receives this name because it has a planet-like orbit, that is, it orbits the center of mass of the system; (2) \textit{S-type}: systems where the planet orbits only one of the binary components. This type of orbit gets its name because it does not orbit the center of mass, but one of the bodies of the system as a satellite \citep{dvorak1982planetary}.

Even with this expressive number of exoplanets confirmed in binary systems, none of them are terrestrial. In the other hand, in the case of short-period binary systems, or close-binaries as they are also known, some works have already shown numerically that terrestrial planets can be formed \citep{lissauer2004terrestrial,quintana2006terrestrial}. In \cite{barbosa2020earth}, the possibility of Earth-sized planets being formed within their habitable zones (HZ) has been investigated. Among all systems studied in this work, it is shown that \textit{Kepler-35} \citep{welsh2012transiting} and \textit{Kepler-38} \citep{orosz2012neptune} systems are the ones with the greatest capacity for this to occur. These results show that the non-existence of a terrestrial planet in binary systems is more associated with the difficulty of discovering these planets in comparison to the gas giants.

The HZ of a system is the region around a star where a terrestrial-mass planet with a $CO_2-H_2O-0N_2$ atmosphere can sustain water in its liquid form on its surface \citep{huang1959occurrence,hart1978evolution,kasting1993habitable,underwood2003evolution,kaltenegger2011exploring}. This broad definition can be extended both to systems with only one star and to multiple star systems. In \cite{haghighipour2013calculating}, a methodology was developed to calculate the limits of this region in binary systems with planets in \textit{P-type} orbits following the model proposed in \cite{kopparapu2013habitable,kopparapu2013erratum}. This method finds the limits of HZ for an Earth-type planet, that is, a gas giant in the same region cannot be considered habitable. This occurs with some planets, such as \textit{Kepler-16b} \citep{doyle2011kepler}, \textit{Kepler-47c,d} \citep{orosz2012kepler,orosz2019discovery} and \textit{Kepler-1647b} \citep{kostov2016kepler}, gas giants within the HZ of their systems.

In \cite{barbosa2020earth} it was investigated the for a set of circumbinary systems (CB) with planets already detected, whether they could form an Earth-type planet within their HZs. However, the \textit{Kepler-1647} system, given its complexity of regions to be studied, has not been investigated. 
The planet of the system, \textit{Kepler-1647b} as is it known, is the most massive CB planet ever discovered by the \textit{Kepler} probe, with $\approx$ 1.5 mass of Jupiter. In addition, it has a very long orbital period ($\approx$ 1100 days), which is also the longest. The planet is around a binary pair consisting of two solar-mass stars that have an orbital period of approximately 11 days \citep{kostov2016kepler}. It has a large semi-major axis of about 2.7 au, which places it within the conservative habitable zone of the system \citep{barbosa2020earth}, being the widest HZ among all CB planetary systems discovered by the \textit{Kepler} probe. 

As the system has the \textit{Kepler-1647b} planet formed, we will consider that the gas disk has already been dissipated in the process of forming this gaseous giant, following the conventional model that terrestrial planets are formed after this phase by accretion of the remaining material of this period \citep{safronov1972evolution,lissauer1993planet}. Thus, the main objective of this work is to study through numerical simulations the possibility of an Earth-size planet be formed inside its HZ. This work was planned in order to make the most complete exploration in terms of initial conditions. As the planet \textit{Kepler-1647b} is close to the center of the HZ, we divided the space of initial conditions into four parts: the regions interior and exterior to the orbit of the planet, the region co-orbital to the planet, and also the satellite region, orbiting around the planet. And in all cases were performed representative sets of simulations that lead to reliable conclusions. The structure of the manuscript is: in section \ref{sec:hz}, the HZ of the Kepler-1647 system was calculated. In section \ref{sec:stability}, we carry out a stability test within the HZ of this system by means of numerical simulations using test particles. In section \ref{sec:planetformation} we study the last stage of planetary formation within the stable regions found in section \ref{sec:stability}. In section \ref{subsec:exomoon}, again through computer simulations, we looked for the stability limits of a moon with a mass equal to that of the Earth around the planet \textit{Kepler-1647b}. Finally, we conclude our work in section \ref{sec:conclusion} emphasizing the main results.

\section{Kepler-1647 habitable zone}
\label{sec:hz}

The planet \textit{Kepler-1647b} has many details that make it stand out in comparison to the other circumbinary planets (CBP) already confirmed, at least so far. One of them is its mass of $\approx$ 1.5 $M_{\text{jup}}$, which puts it as the most massive. Another odd detail is its longest orbital period of all, with a semi-major axis of $\approx$ 2.72 $au$ its orbital period 1107.59 days \citep{kostov2016kepler}. The system's stars are also the most massive among CB systems with planets detected by the \textit{Kepler} probe \citep{borucki2010kepler} and its \textit{K2} upgrade \citep{howell2014k2}. Both have around a solar mass and complete a period in 11 days with a moderate eccentricity ($ e_ {\text{bin}} $ = 0.16). For more details of the system check Table \ref{tab:binarydata}. The luminosity of the stars shown in this table were calculated using
\begin{equation}
    \centering
    \label{eq:lumino}
    \frac{L}{L_\odot}=\left(    \frac{R}{R_\odot}   \right)^2\left(    \frac{T}{T_\odot}   \right)^4,
\end{equation}
where L, T, and R represent the luminosity, effective temperature and radius of the star to be calculated, respectively and $L_\odot$, $R_\odot$, and $T_\odot$ are the luminosity, radius, and effective temperature of the Sun \citep{duric2004advanced}.

\begin{table}
\caption{Mutual inclination refers to the inclination between the planet and the binary stars. * the star's luminosity was calculed using equation \ref{eq:lumino}.}
\label{tab:binarydata}
\centering
\begin{tabular}{lcc}
\hline
\hline
\multicolumn{3}{l}{Binary star data}                  \\
\hline
Primary mass, $M_A$           & 1.221  & $M_{\odot}$ \\
Secondary mass, $M_B$           & 0.968  & $M_{\odot}$ \\
Primary radius, $R_A$           & 1.790  & $R_{\odot}$ \\
Secondary radius, $R_B$           & 0.966  & $R_{\odot}$ \\
Primary temperature, $T_A$           & 6210   & K                           \\
Secondary temperature, $T_B$           & 5770   & K                           \\
Primary luminosity*, $L_A$           &     4.269   & $L_{\odot}$ \\
Secondary luminosity*, $L_B$           &    0.927    & $L_{\odot}$ \\
Orbital period & 11.259 & days                        \\
Semimajor axis, $a_{\text{bin}}$     & 0.128  & au                          \\
Eccentricity, $e_{\text{bin}}$     & 0.160  &            \\   
\hline
\multicolumn{3}{l}{Planet data}            \\
\hline
Orbital period     & 1,107.592 & days       \\
Semimajor axis, $a_p$               & 2.721    & au         \\
Eccentricity, $e_p$               & 0.058    &            \\
Mutual inclination & 2.985    & °          \\
Mass               & 1.520    & $M_{\text{jup}}$ \\
Reference   & \multicolumn{2}{c}{\cite{kostov2016kepler}}   \\
\hline
\end{tabular}
\end{table}

 In the last decade, together with the confirmation of planets in binary systems, many works have endeavored to know the conditions and limits of the HZ of these systems \citep{kane2012habitable, eggl2012analytic, quarles2012habitability, mason2013rotational, liu2013find, eggl2013circumstellar, haghighipour2013calculating,kaltenegger2013calculating,muller2014calculating, mason2015circumbinary, zuluaga2016constraining}.

\begin{figure}
    \centering
	\includegraphics[width=\columnwidth]{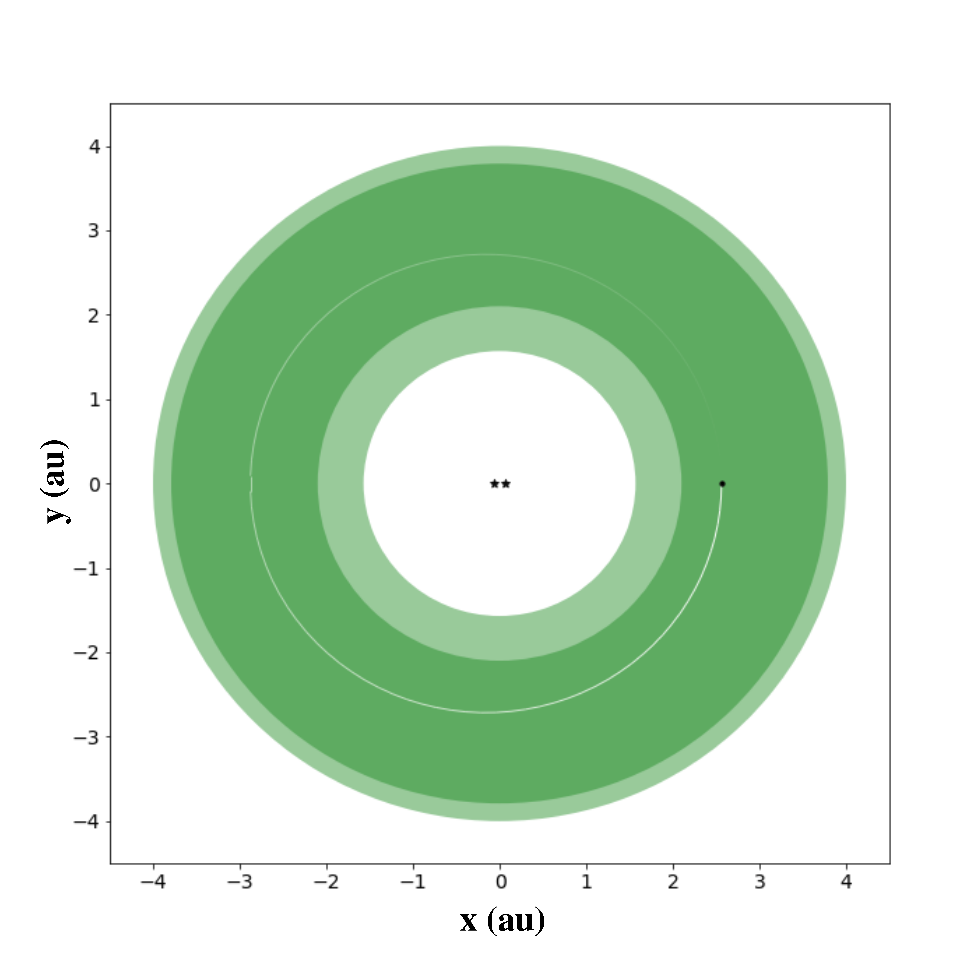}
    \caption{Orbital plane of the binary pair of the \textit{Kepler-1647} system showing its HZ limits. The extended and conservative HZ \citep{kopparapu2013habitable, kopparapu2013erratum} are represented in light and dark green. The figure is centered on the center of mass of bodies (stars and planet). The black stars in the center represent the stars of the system and the point with the white trail the planet \textit{Kepler-1647b.}}
    \label{fig:hz1647}
\end{figure}

In this work, to calculate the HZ limits of the \textit{Kepler-1647} system, we used a model proposed by \cite{haghighipour2013calculating}. Its work generalizes the calculation of the HZ limits to an Earth-type planet defined in \cite{kopparapu2013habitable,kopparapu2013erratum} for binary systems with planets in P-type orbits. The Figure \ref{fig:hz1647} shows the HZ of the system and its limits can be seen more precisely in the Table \ref{tab:hzlimits1647}. In this Figure, the stars are in the center and the planet is orbiting within the HZ with the trace of its orbit in white. It is important to remember that even though it is within the HZ, the planet cannot be considered habitable due to the fact that these limits are calculated for an Earth-type planet.

\begin{table}
\caption{Kepler-1647 HZ boundaries.}
\label{tab:hzlimits1647}
\centering
\begin{tabular}{ccccc}
\hline
\hline
\multirow{2}{*}{} & \multicolumn{2}{c}{Conservative HZ} & \multicolumn{2}{c}{Extended HZ} \\ \cline{2-5}
                        & Inner (au)    & Outer (au)    & Inner (au)   & Outer (au)   \\
Kepler-1647             & 2.10          & 3.79          & 1.57         & 4.00        \\
\hline
\end{tabular}
\end{table}

\section{HZ stability test}
\label{sec:stability}
Before performing any numerical simulation of planetary formation, we need to investigate the stability of the system. The stability we are referring to is the location where particles survive within the HZ for more than one million years, that is, they are not ejected out of that region neither collide with the stars or the planet. In the work of \cite{quarles2018stability}, the authors studied the limits of stability of the CB planets confirmed by the \textit{Kepler} mission. They showed that the greater the eccentricity of the binary pair, the farther from the stars is the inner stability limit. In the case of the \textit{Kepler-1647} system, they have shown that the $ a_c $ limit is 0.3497 au. The same is shown in \cite{pichardo2005circumstellar,10.1111/j.1365-2966.2008.13916.x}, looking for \textit{"invariant loops"} using test particles, it is shown that the size of the stable disc around the stars decreases according to the increase of the eccentricity of the binary stars.

Bearing in mind that the planet Kepler-1647b has a semi-major axis equal to 2.721 au, there is a large region between this limit of internal stability up to the planet. For this reason, \cite{kostov2016kepler} also checked the conditions for a hypothetical planet (with the same mass of the Kepler-1647b) to be stable in that region using Mean Exponential Growth factor of Nearby Orbits (MEGNO) formalism \citep{gozdziewski2001global,cincotta2003phase,hinse2015predicting}. The authors have found that there is a stable region from $ \approx $ 0.5 to $ \approx $ 2.0 au.

In our work, as we are interested in exploring the formation of an Earth-sized planet within the HZ, we need to investigate stability throughout this region. Our goal here is to find the limits of stability so that later we can carry out the simulations of planetary formation. For this, we performed numerical simulations using MINOR-MERCURY package, an adaptation of the MERCURY package \citep{chambers1999hybrid}, with option for close binaries made by us following \cite{chambers2002symplectic}, that was extensively tested in \cite{barbosa2020earth}.

\subsection{Initial conditions}
\label{subsec:HZic}

As we are interested in exploring the formation of an Earth-sized planet within the HZ, we will investigate the stability of the system within this region +20\% (Table \ref{tab:stal}). For this, we used massless particles that interact gravitationally only with the host planet and the stars of the system, that is, without mutual interaction between them. Particles were randomly distributed between the internal and external limits with an extra margin of 20\% of the entire width of the extended HZ. For more details, see Table \ref{tab:stal}. 

The initial eccentricity of each particle varied between 0.0 and 0.01 and the initial inclination between $10^{- 5}$ and $10^{- 4}$ degrees. We despise the fact that the planet is within HZ to study the stability limit around it and also to investigate possible stability in its co-orbital region. Thus, in our simulations are present the stars, the host planet of the Kepler-1647 system with all its real data that can be checked in Table \ref{tab:binarydata}, and the test particles. Ejection is considered to be particles exceeding 10 au from the center of mass of the system or exceeding the interior stability limit ($a_c$ = 0.3497 au) shown in \cite{quarles2018stability}. The length of simulation time is one million years.

\begin{table}
\caption{Initial conditions of the massless particles for stability test.}
\label{tab:stal}
\centering
\begin{tabular}{lcc}
\hline
\hline
Number of particles                         & 10,000 &                \\
Inner limit (with -20\%) & 1.081  & au             \\
Outer limit (with +20\%) & 4.486  & au             \\
Inclination & $10^{- 5}-10^{- 4}$ & degree \\
Eccentricity & 0.0-0.01 & \\
\hline
\end{tabular}
\end{table}

\subsection{Results}
\label{subsec:stallresult}

\begin{figure*}
    \centering
	\includegraphics[scale=0.7]{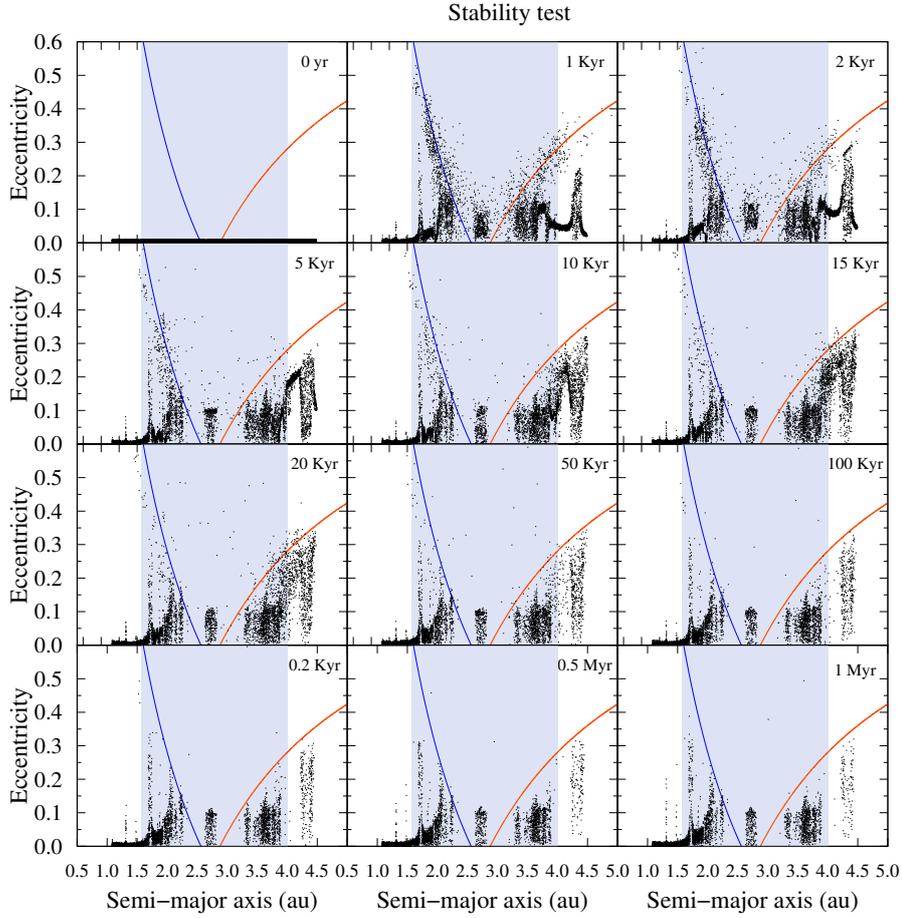}
    \caption{Snapshots in time of the dynamic evolution of test particles in the HZ. The horizontal and vertical axes are the semi-major axis and the eccentricity, respectively. The black dots are the particles and red lines represent the apocentre of the host planets of the system in function of the pericentre of the particle, given by $a = [a_p (1 + e_p )]/(1 - e)$, and the blue lines represent the pericentre of the planet as a function of the apocentre of the particles, given by $a = [a_p (1 - e_p )]/(1 + e)$. Where $a_p$ and $e_p$ are the semi-major axis and eccentricity of the planet respectively. The shaded region in blue represents HZ of the system.}
    \label{fig:stal1647}
\end{figure*}

Figure \ref{fig:stal1647} shows the dynamic evolution of particles over time. As expected, we can note that the host planet, with more than 1.5 $M_{\text{jup}}$, produces a lot of disturbance in the disk. Its presence makes particles close to it to have a considerable increase in eccentricity in the first thousands of years of simulation. This increase causes the particles to cross the planet's orbit, colliding with it or being ejected from the system. From the initial 10,000 particles, 55\% are ejected and 0.5\% collide with the planet. The outermost regions of the disk have the particles with the largest eccentricity.

\begin{figure*}
    \centering
	\includegraphics[scale=0.30]{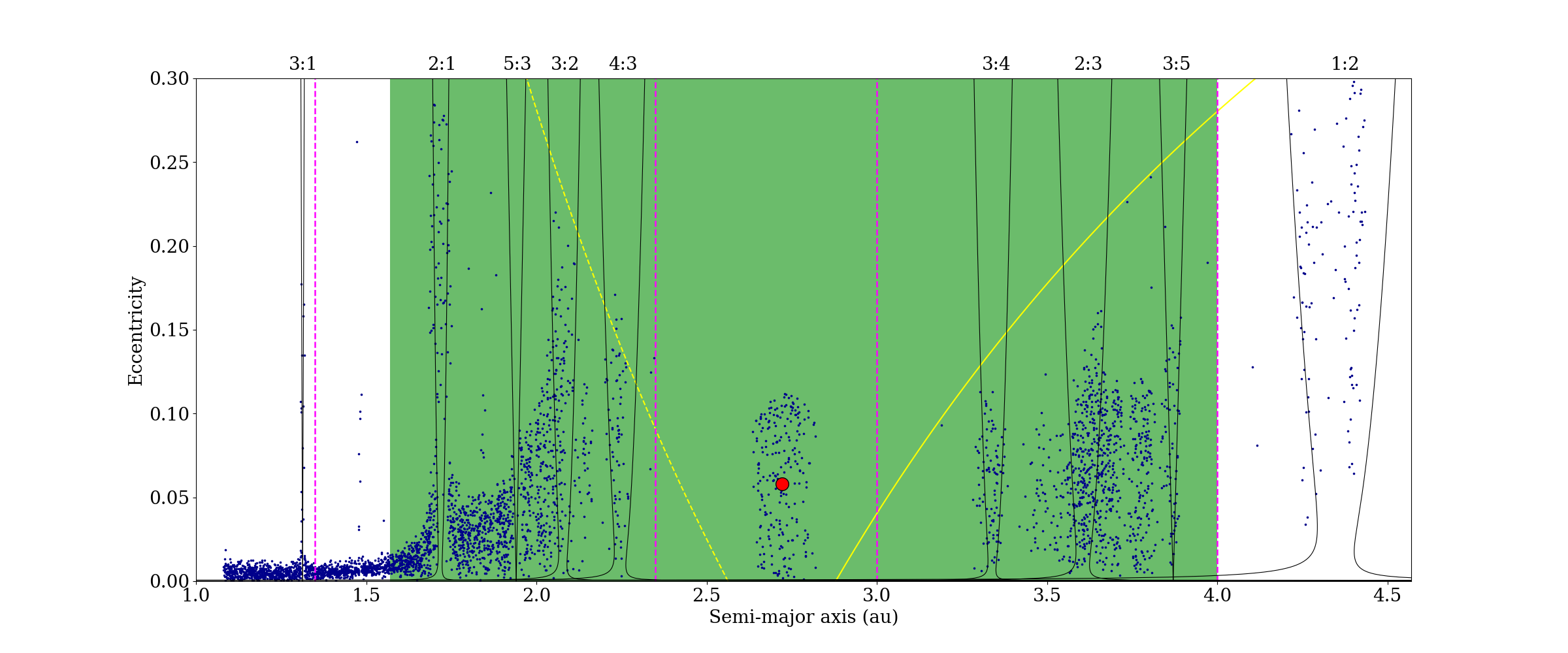}
    \caption{The final state of the test particles and the planet in one million years showing the eccentricity versus the semi-major axis. The blue dots represent the particles and the red dot the planet. Yellow solid line represents the apocentre of the host planet of the system in function of the pericentre of the particles, given by $a = [a_p (1 + e_p )]/(1 - e)$, and the yellow dashed line represents the pericentre of the planet as a function of the apocentre of the particles, given by $a = [a_p (1 - e_p )]/(1 + e)$, where $a_p$ and $e_p$ are the semi-major axis and eccentricity of the planet, respectively. The shaded region in green denotes the HZ. The dashed magenta lines indicate the initial position limits of particles for the planetary formation simulations. The black lines denote the maximum libration zones as a function of semi-major axis and eccentricity for a selection of resonances.The nominal resonance locations are indicated on the top the plot.}
    \label{fig:zoon}
\end{figure*}

At the end of the simulation (Figures \ref{fig:stal1647} and \ref{fig:zoon}), we have that it can be divided into three stable sub-regions (at least), one internal to the planet, one coorbital and one external to it. In this way, there are three regions where we can explore the planetary formation. The stability limits of the co-orbital region needed to be studied in more detail, and this will be discussed in the next subsection. However, the limits of the internal and external regions of the planet can be seen in Figure \ref{fig:zoon}. The dashed magenta lines show the limits where they are and will be used to distribute the disk of matter for the study of planetary formation.

\subsection{Co-orbital stability}

Particles in the co-orbital region share the same orbit with the CB planet, close to the stable Lagrangian equilibrium points $L_4$ and $L_5$. We note in Figure \ref{fig:zoon} a considerable volume of remaining particles in this region. Since these orbits are extremely sensitive and unknown in binary systems, we are going to show here the study of this particular region in more detail.

 As these regions close to the equilibrium points are symmetrical, we perform this test only around the point $L_4$ and the results found for it will be similar for $L_5$. 5,000 particles were distributed between 2.22-3.22 au, 
 corresponding to twice the largest width of the horseshoe orbit given by
 \begin{equation}
     \centering
     \Delta_{\text{horse}}=\mu^{\frac{1}{3}}a_p,
 \end{equation}
 \citep{dermott1981dynamics} where $\mu$ is the relative mass and $a_p$ the semi-major axis of the \textit{Kepler-1647b}. The eccentricity and inclination used was the same of the planet co-orbital to them. We use these values because the secular perturbations caused by the planet in the particles forces them to have the same eccentricity and inclination as itself \citep{murray1999solar}. Check Table \ref{tab:param_coo} for more details on initial conditions of the simulation. The particles only interact gravitationally with the stars and the planet there is no mutual interaction among them. The length of integration time is also one million years.

\begin{figure*}
    \centering
	\includegraphics[scale=0.5]{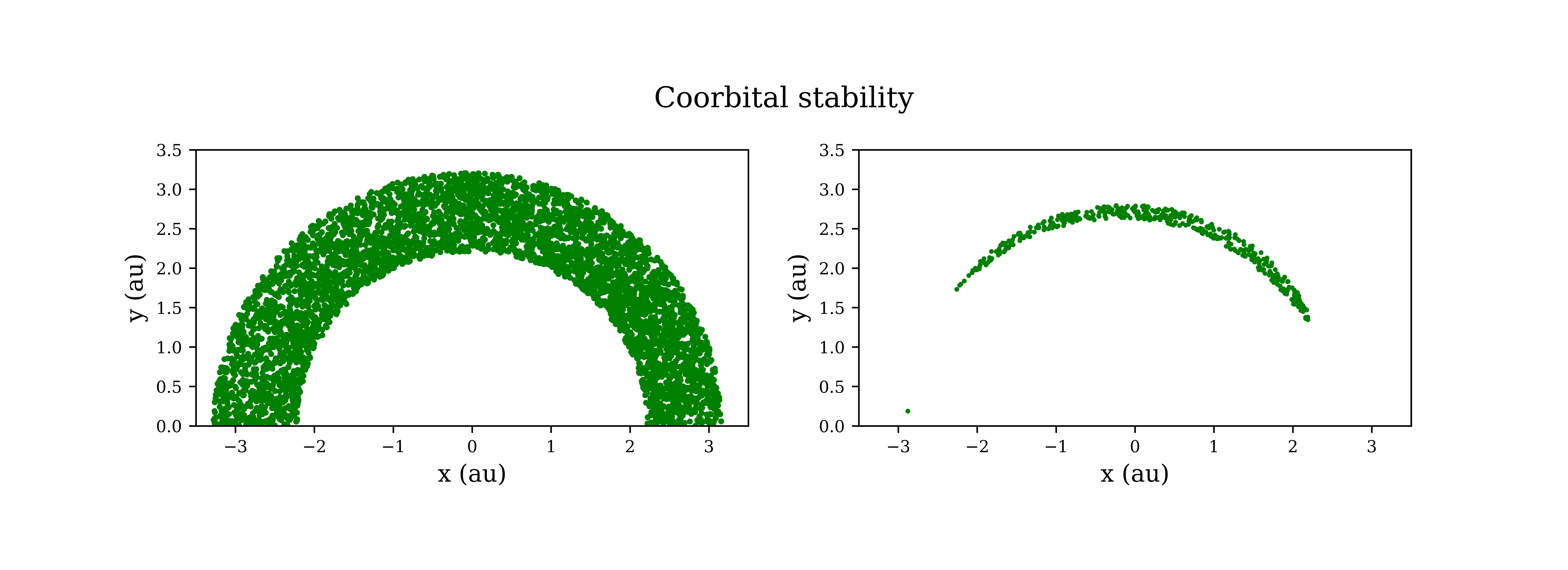}
    \caption{The two sub-figures show the positions of the particles in Cartesian x and y coordinates. In (a) all 5,000 particles are present in their initial positions and in (b) the initial positions of the surviving particles.}
    \label{fig:coo_stall}
\end{figure*}

Figure \ref{fig:coo_stall} shows the result of the simulation. From the initial 5,000 particles, 472 survived. Is shown in \textit{(a)} the initial position of the particles in cartesian $x$ and $y$ coordinates, and in \textit{(b)}, the initial condition of the surviving particles at the end of the integration. This result shows that even though the system is binary, there is a considerable stable region. We will use the orbital elements of the surviving particles to test the planetary formation in this region.

\begin{table}
\centering
\caption{Parameters of the stability test in the co-orbital region.}
\label{tab:param_coo}
\begin{tabular}{lcc}
\cline{2-3}
\cline{2-3}
                                          & Values & Unit \\ \hline
Number of particles                       & 5,000  & un   \\
\multicolumn{1}{c}{Semi-major axis (min)} & 2.220  & au   \\
\multicolumn{1}{c}{Semi-major axis (max)} & 3.220  & au   \\
Eccentricity                              & 0.058  &      \\
Inclination                               & 2.986  & °    \\
Mean anomaly (min)                        & 0      & °    \\
Mean anomaly (max)                        & 180    & °    \\ \hline
\end{tabular}
\end{table}

\section{Planetary formation}
\label{sec:planetformation}

The stability test provided us with three different scenarios to explore the planetary formation in this system. Being an internal region, a coorbital and one external to the host planet. Therefore, we set up a set of simulations for these three cases.

\subsection{Co-orbital region}

Studies of the formation of terrestrial planets in co-orbital regions have already been explored in one-star systems. In \cite{beauge2007co}, the authors numerically investigated different scenarios for the formation of terrestrial planets using a hypothetical system with a planet similar to Jupiter orbiting a Sun like star. In their simulations an N-body integrator, the authors showed that it is not possible to form a Trojan planet with a mass larger than 0.6 $M_\oplus$. With a different context, \cite{izidoro2010co} also numerically explored the formation of smaller bodies in the co-orbital region of a satellite orbiting a planet. The work of \cite{10.1093/mnras/staa1727} studied the structure of co-orbital stable regions for a wide range of mass ratio systems and provide empirical equations to describe them.

We began our study by assuming that particles with mass are trapped around 
around what would be the $L_4$ point of Lagrangian equilibrium in a Solar type system. As we can see in the stability test shown in the previous section, there is a stable portion sharing the orbit with the host planet, like the Trojans in the case of the solar system. Thus, taking into account the stability found, we studied the possibility of a larger body being formed in this region. As it is a very sensitive region from a dynamic point of view, we explore different mass values around this region.

\subsubsection{Initial conditions}

In order to explore this region to the full, we used eight different total mass values, being 0.4, 0.6, 0.8, 1.0, 1.2, 1.4, 1.6 and 1.8 $M_{\oplus}$. In all of these cases, we used the initial positions of the surviving particles from the stability test in the co-orbital region (see Figure \ref{fig:coo_stall}).
The eccentricities and inclinations of the particles were also kept the same used in the stability test of the co-orbital region, which are, the same as the host planet. However, as we are now interested in bodies that increase mass and can become planets, all bodies interact gravitationally with each other. In the simulations we have the two stars, the host planet, and the particles.

\begin{figure}
    \centering
	\includegraphics[width=\columnwidth]{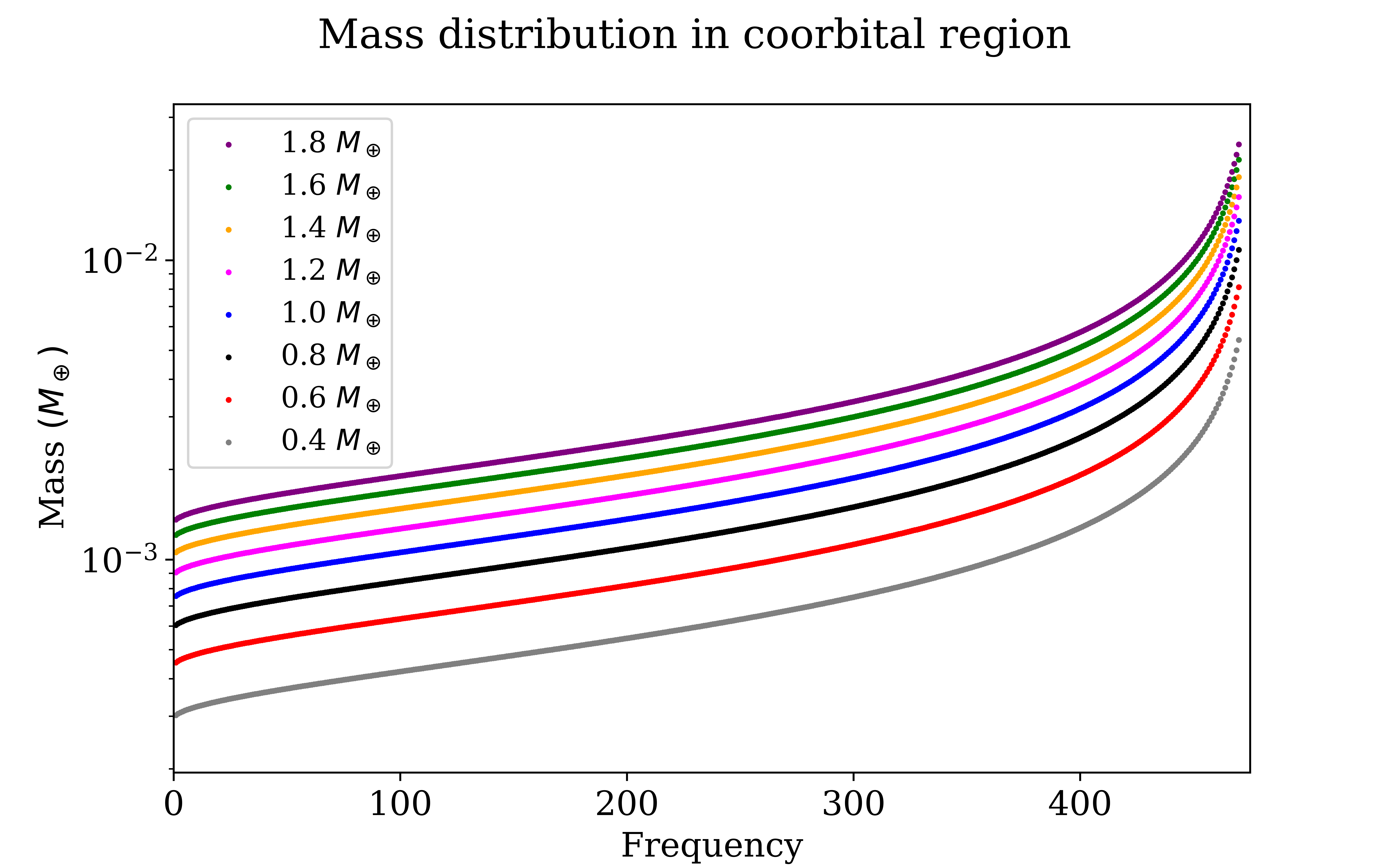}
    \caption{Mass distribution of particles for different values of total mass.}
    \label{fig:dist_same}
\end{figure}

For each of these total mass values, we performed 10 different simulations, resulting in 80 simulations, where we randomly vary their masses. With the eight different values of total mass, we distributed the particle's masses following a power-law. In all distributions there is a large amount of less massive particles than particles with a higher mass, see Figure \ref{fig:dist_same}.

The power-law used was,
\begin{equation}
\centering
m_p(\delta_i)=MT_j(\alpha+\beta\sqrt{\delta_i})^\gamma,
\label{eq:dist_func}
\end{equation}
where $m_{p}$, $MT_j$ $(j=1,2,...,8)$ and $\delta_i$ ($i=1, 2, ..., 471$) represent the individual mass of each particle, the total mass and the frequency of distribution respectively. The values of $\alpha$, $\beta$ and $\gamma$ are constant coefficients equal to 0.60318E+14, -0.27616E+13 and -0.59436 found numerically. The simulations were carried out by 200 Myr.

\subsubsection{Results}

As mentioned in previous sections, we performed numerical simulations using as initial conditions the initial positions of the surviving particles in the co-orbital region stability test. Ten simulations were performed for each total mass value, totaling 80 simulations. Next, we will show the results of each of these simulations.

As an analysis of the results, we defined as an Earth type planet being a body with a mass close to that of Earth and within HZ.
A first analysis of our results, show that in none of the simulations an Earth-type was formed. Table \ref{tab:coo_result} shows the results found over 200 million years of the simulations with total masses from 0.4 to 1.8 Earth masses. The first column indicates the simulation id, composed of XX-Y, XX being the total mass and Y the number of the simulation, which varies from 1 to 10. 

\begin{table*}
\centering
\caption{The columns show from left to right the identification number of the simulation, the number of surviving bodies, the surviving particle with the greatest mass and, the semi-major axis, the eccentricity and inclination of this particle and its mass.}
\label{tab:coo_result}
\begin{tabular}{ccccccccccccccc}
\hline
\hline
\multicolumn{15}{c}{Results of simulations of planetary formation in the co-orbital region} \\ \hline
Sim    & SP & Particle & $a_f$  & $e$    & $i$        & Mass        &  & Sim    & SP & Particle & $a_f$  & $e$     & $i$        & Mass        \\ \cline{4-4} \cline{6-7} \cline{12-12} \cline{14-15} 
       &    &          & $(au)$ &        & ($^\circ$) & ($M_\odot$) &  &        &    &          & $(au)$ &         & ($^\circ$) & ($M_\odot$) \\ \hline
0.4-1  & 2  & part4876 & 2.7843 & 0.0693 & 2.7694     & 7.6841e-4  &  & 1.2-1  & 2  & part1973 & 2.7158 & 0.0897  & 2.6444     & 7.0423e-3  \\
0.4-2  & 5  & part4541 & 2.6778 & 0.1353 & 2.8524     & 1.7559e-3  &  & 1.2-2  & 2  & part901  & 2.7263 & 0.1269  & 4.4843     & 2.0590e-3  \\
0.4-3  & 2  & part1109 & 2.7547 & 0.0688 & 3.5176     & 1.2799e-2  &  & 1.2-3  & 2  & part4876 & 2.6747 & 0.1215  & 3.9838     & 6.6946e-3  \\
0.4-4  & 3  & part452  & 2.6908 & 0.0794 & 2.8596     & 8.2690e-3  &  & 1.2-4  & 3  & part2747 & 2.6887 & 0.0946  & 2.5085     & 4.0041e-2  \\
0.4-5  & 2  & part1734 & 2.7150 & 0.0643 & 2.9806     & 1.0349e-2  &  & 1.2-5  & 2  & part3443 & 2.6745 & 0.0633  & 3.1422     & 2.6919e-2  \\
0.4-6  & 2  & part3570 & 2.7107 & 0.0136 & 1.8376     & 7.4712e-3  &  & 1.2-6  & 3  & part4876 & 2.7554 & 0.0499  & 4.0617     & 1.9971e-2  \\
0.4-7  & 3  & part3073 & 2.7660 & 0.0895 & 3.1394     & 8.3442e-3  &  & 1.2-7  & 3  & part2847 & 2.7431 & 0.0583  & 4.1182     & 2.7428e-3  \\
0.4-8  & 2  & part2434 & 2.7509 & 0.0516 & 1.8793     & 4.5088e-3  &  & 1.2-8  & 3  & part890  & 2.6683 & 0.0770  & 4.1952     & 1.2612e-2  \\
0.4-9  & 2  & part3124 & 2.7524 & 0.0334 & 3.4045     & 8.0364e-3  &  & 1.2-9  & 2  & part1942 & 2.6940 & 0.0241  & 3.4894     & 1.2600e-2  \\
0.4-10 & 2  & part3224 & 2.7696 & 0.0352 & 2.8087     & 2.0903e-2  &  & 1.2-10 & 2  & part2128 & 2.7333 & 0.0501  & 3.3821     & 1.2477e-2  \\ \cline{1-7} \cline{9-15} 
0.6-1  & 2  & part2922 & 2.7097 & 0.0958 & 2.2859     & 6.2828e-3  &  & 1.4-1  & 3  & pt2192   & 2.7348 & 0.0707  & 6.8415     & 2.1837e-2  \\
0.6-2  & 5  & part805  & 2.6854 & 0.0666 & 3.4304     & 1.5242e-3  &  & 1.4-2  & 3  & part1180 & 2.6942 & 0.1053  & 3.7177     & 1.7488e-2  \\
0.6-3  & 3  & part2921 & 2.7064 & 0.1253 & 3.3277     & 4.5469e-3  &  & 1.4-3  & 3  & part1761 & 2.7285 & 0.0476  & 3.8518     & 1.3281e-2  \\
0.6-4  & 2  & part4372 & 2.7455 & 0.0599 & 3.5614     & 1.6173e-2  &  & 1.4-4  & 2  & part720  & 2.7154 & 0.0408  & 1.9427     & 2.6899e-2  \\
0.6-5  & 2  & part2704 & 2.6696 & 0.0695 & 2.7640     & 2.3414e-2  &  & 1.4-5  & 3  & part1251 & 2.7901 & 0.05868 & 2.6928     & 2.8419e-2  \\
0.6-6  & 3  & part3222 & 2.7483 & 0.0445 & 2.118      & 8.9929e-3  &  & 1.4-6  & 3  & part3813 & 2.6856 & 0.04263 & 2.4857     & 2.6621e-3  \\
0.6-7  & 3  & part2929 & 2.7227 & 0.0509 & 2.5468     & 8.6247e-3  &  & 1.4-7  & 3  & part3321 & 2.7266 & 0.0544  & 3.9329     & 4.2609e-2  \\
0.6-8  & 2  & part3542 & 2.6896 & 0.1053 & 3.4187     & 5.6994e-3  &  & 1.4-8  & 1  & part2718 & 2.7370 & 0.0667  & 2.9659     & 3.0811e-2  \\
0.6-9  & 2  & part1009 & 2.6966 & 0.0934 & 3.3951     & 9.5336e-3  &  & 1.4-9  & 2  & part1238 & 2.7256 & 0.0608  & 4.0384     & 1.2338e-2  \\
0.6-10 & 3  & part261  & 2.7113 & 0.1026 & 2.9009     & 4.4257e-3  &  & 1.4-10 & 2  & part1694 & 2.6729 & 0.0468  & 2.6681     & 6.0304e-3  \\ \cline{1-7} \cline{9-15} 
0.8-1  & 2  & part3035 & 2.7007 & 0.0772 & 3.0358     & 1.4528e-2  &  & 1.6-1  & 2  & part1399 & 2.7238 & 0.0524  & 2.0200     & 3.3219e-2  \\
0.8-2  & 3  & part4173 & 2.7853 & 0.0878 & 3.5697     & 1.2905e-2  &  & 1.6-2  & 2  & part2070 & 2.7340 & 0.0433  & 3.9092     & 8.6235e-3  \\
0.8-3  & 2  & part3375 & 2.6840 & 0.0064 & 2.8048     & 1.0228e-2  &  & 1.6-3  & 4  & part2805 & 2.6834 & 0.0835  & 0.7809     & 3.8656e-3  \\
0.8-4  & 5  & part1055 & 2.6944 & 0.1293 & 3.5695     & 2.4642e-3  &  & 1.6-4  & 3  & part560  & 2.8057 & 0.0660  & 4.0955     & 7.6440e-3  \\
0.8-5  & 5  & part1734 & 2.7432 & 0.0183 & 3.1201     & 1.2049e-2  &  & 1.6-5  & 2  & part3756 & 2.7329 & 0.0564  & 2.9696     & 2.6583e-2  \\
0.8-6  & 2  & part1966 & 2.7460 & 0.0452 & 3.7021     & 1.0597e-2  &  & 1.6-6  & 3  & part2320 & 2.7278 & 0.0725  & 2.8423     & 1.3899e-2  \\
0.8-7  & 3  & part1655 & 2.6967 & 0.0514 & 3.8889     & 3.6993e-3  &  & 1.6-7  & 1  & part3280 & 2.6796 & 0.0324  & 3.0362     & 1.3260e-2  \\
0.8-8  & 2  & part3331 & 2.7358 & 0.0384 & 2.5727     & 5.1276e-3  &  & 1.6-8  & 2  & part158  & 2.7225 & 0.1061  & 1.5238     & 1.9593e-2  \\
0.8-9  & 2  & part2896 & 2.6666 & 0.0447 & 3.6505     & 4.6755e-3  &  & 1.6-9  & 2  & part2558 & 2.7816 & 0.0576  & 2.1428     & 2.0084e-2  \\
0.8-10 & 3  & part1753 & 2.7214 & 0.0684 & 2.6985     & 1.1620e-2  &  & 1.6-10 & 3  & part471  & 2.7412 & 0.0687  & 2.3680     & 2.5096e-2  \\ \cline{1-7} \cline{9-15} 
1.0-1  & 3  & part4815 & 2.7005 & 0.0646 & 3.2192     & 4.9347e-3  &  & 1.8-1  & 3  & part2998 & 2.7643 & 0.0804  & 2.1566     & 2.4359e-3  \\
1.0-2  & 2  & part4557 & 2.7178 & 0.0815 & 1.8794     & 1.3740e-2  &  & 1.8-2  & 2  & part733  & 2.7455 & 0.0743  & 3.6756     & 4.2858e-3  \\
1.0-3  & 3  & part1109 & 2.7128 & 0.0165 & 2.1330     & 5.9580e-3  &  & 1.8-3  & 3  & part541  & 2.7006 & 0.0626  & 2.5949     & 3.4212e-2  \\
1.0-4  & 3  & part2095 & 2.6467 & 0.0518 & 2.5262     & 3.1756e-3  &  & 1.8-4  & 2  & part3443 & 2.6908 & 0.1019  & 1.0418     & 3.0506e-3  \\
1.0-5  & 2  & part2704 & 2.7453 & 0.0348 & 3.2974     & 1.6286e-2  &  & 1.8-5  & 2  & part1646 & 2.7357 & 0.1122  & 2.5803     & 6.0330e-2  \\
1.0-6  & 2  & part3222 & 2.6792 & 0.0590 & 2.9855     & 8.9929e-3  &  & 1.8-6  & 2  & part4042 & 2.7477 & 0.1008  & 3.3634     & 2.1837e-2  \\
1.0-7  & 2  & part2921 & 2.7388 & 0.0986 & 2.8586     & 5.8992e-3  &  & 1.8-7  & 3  & part4550 & 2.7310 & 0.0095  & 3.1216     & 3.6401e-2  \\
1.0-8  & 3  & part4500 & 2.6878 & 0.1181 & 4.3627     & 1.3708e-3  &  & 1.8-8  & 3  & part4098 & 2.7412 & 0.0544  & 3.1086     & 7.5901e-2  \\
1.0-9  & 3  & part582  & 2.7952 & 0.0588 & 2.9855     & 8.9929e-3  &  & 1.8-9  & 2  & part1399 & 2.7198 & 0.0515  & 1.1901     & 1.7153e-2  \\
1.0-10 & 2  & part2138 & 2.6847 & 0.0585 & 3.2003     & 9.6590e-3  &  & 1.8-10 & 1  & part4525 & 2.7431 & 0.0916  & 3.5973     & 8.4359e-3  \\ 
\hline
\hline
\end{tabular}
\end{table*}

The region co-orbital to a planet is recognized as a sensitive region from stability in Solar type systems. In the case of binary systems, this region is even more sensitive as we can see from Table \ref{tab:coo_result}, where few bodies survive the simulation and there is practically no mass addition. On the other hand, we can note that in all cases with different values of the total mass, bodies survived. This shows us that CB systems can have Trojans asteroids as in the case of the Solar system.

\subsection{Inner and outer regions of the planet}

As shown in Figure \ref{fig:zoon}, in addition to the co-orbital region, two others have some stability, an internal and an external region. In order to explore the terrestrial planetary formation in these regions, we performed 40 simulations with different initial conditions. In all of them, we again use our numerical package adapted with the option for close-binary systems. The largest and which has the largest number of particles that survive the stability test is the internal one. Besides, the beginning of the internal part demonstrates that the HZ surviving particles have less eccentricity than the outer region. In all simulations, the binary pair and the giant planet of the system will be present in addition to the circumbinary disk of matter. The data of the host bodies of the system are the real ones found in the literature and can be checked in Table \ref{tab:binarydata}.

\subsubsection{Initial conditions}

\begin{figure}
    \centering
	\includegraphics[width=\columnwidth]{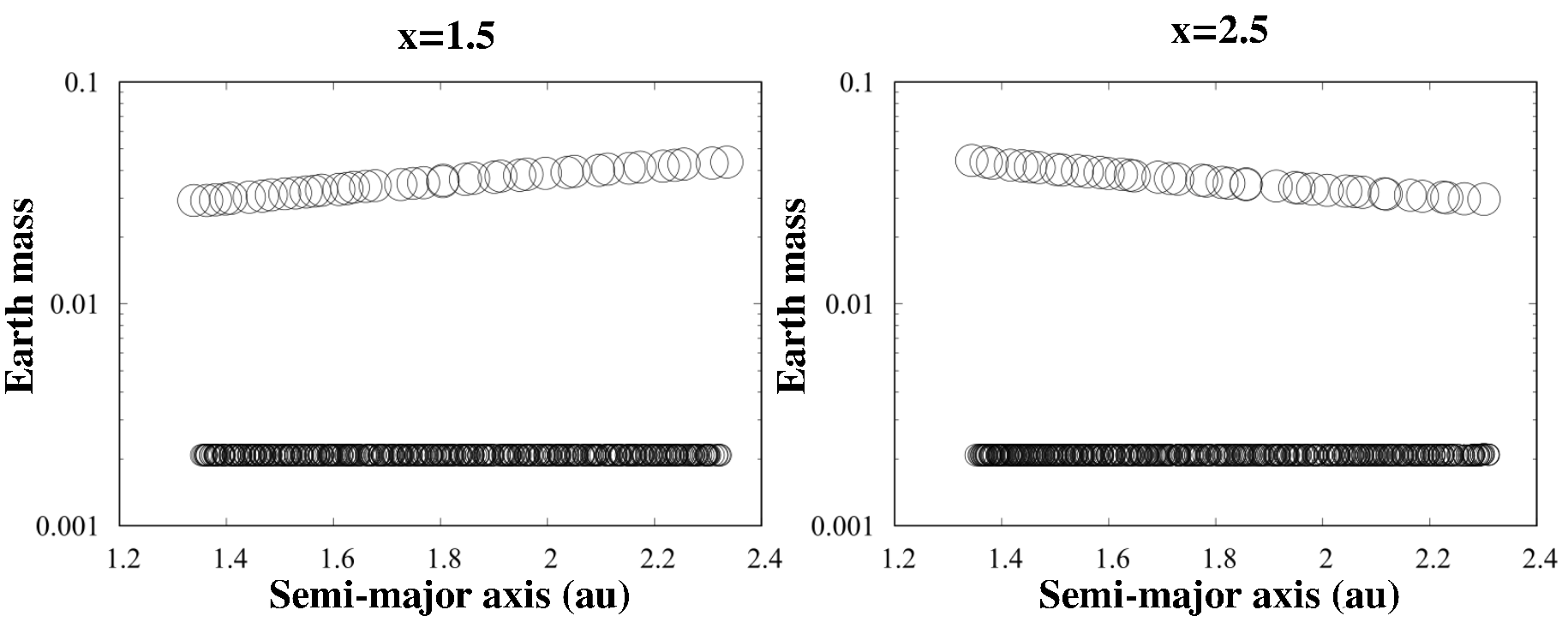}
    \caption{Initial distribution of embryos and planetesimals in the internal stable region. On the left the mass distribution is shown using the coefficient $x = 1.5$ while on the right $x = 2.5$.}
\label{fig:ini_form_z1}
\end{figure}

\begin{figure}
    \centering
	\includegraphics[width=\columnwidth]{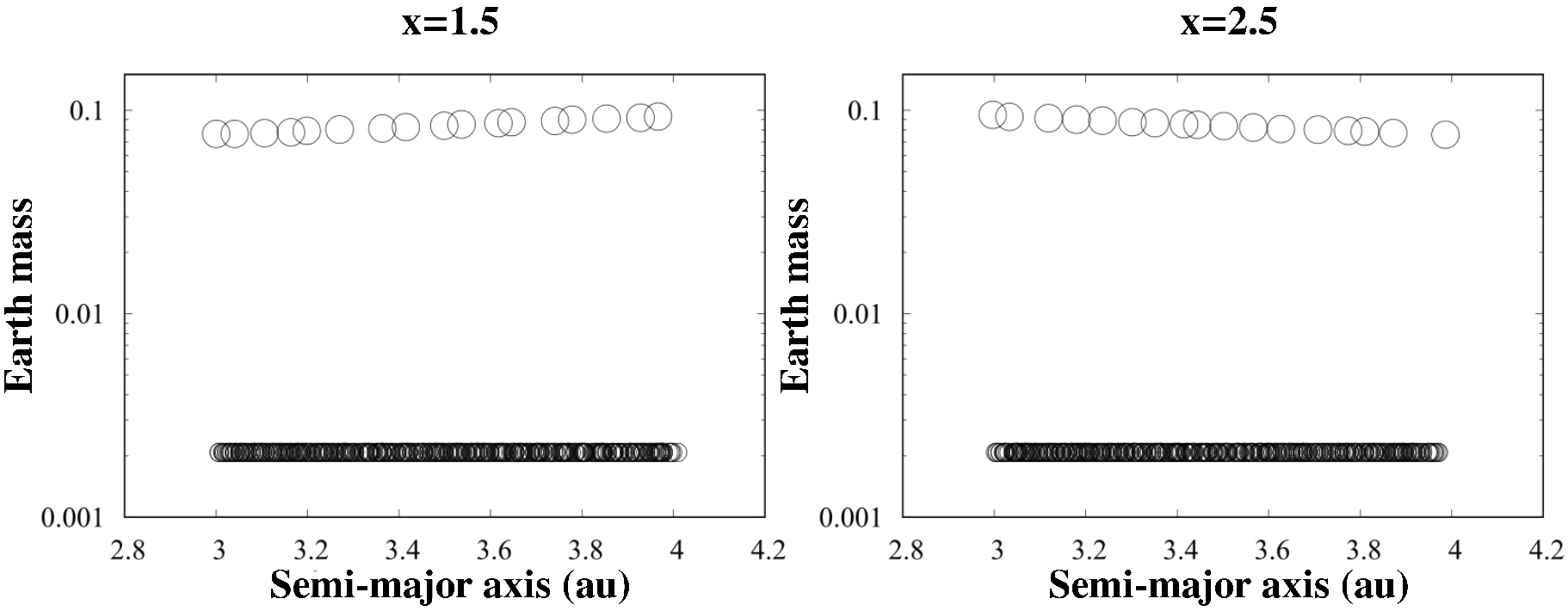}
    \caption{Initial distribution of embryos and planetesimals in the external stable region. On the left the mass distribution is shown using the coefficient $x = 1.5$ while on the right $x = 2.5$.}
    \label{fig:ini_form_z2}
\end{figure}

To explore the planetary formation in these two regions, we used a protoplanetary disks with bi-modal mass distribution composed of planetesimals and planetary embryos, resulting from the runaway and oligarchic growths, as shown by \cite{kokubo1998oligarchic,kokubo2000formation}. Following this context, the mass of the total planetesimals represents 40\% of the total mass of the disk while the sum of the masses of the embryos the other 60\%. The major role of planetesimals is to provide dynamic friction to dump the variation of eccentricities and inclinations of planetary embryos \citep{o2006terrestrial,morishima2008formation}. The internal disk has a width of 1 $au$ and extends from 1.35 to 2.35 $au$ and the external disk, also with 1 $au$ wide, extends from 3 to 4 $au$. The individual mass of the planetesimals is $\approx$ 0.0021 $M_\oplus$ and they are distributed with a surface density profile of $\Sigma_1 r^{-x}$, where $r$ is the radial distance and $\Sigma_1$ is a solid surface density adjusted to fill a total mass of 2.5 $M_\oplus$ within the disk. The mass of each embryo scales following
\begin{equation}
    M_{emb} \sim \Delta^{3/2} r^{3/2(2-x)},
\end{equation}
\citep{kokubo2002formation, raymond2005terrestrial}, where $x$ is a free parameter of the surface density profile and $\Delta$ the mutual separation in Hill radius. We set $\Delta$ randomly by $5-10$ Hill radii \citep{kokubo2000formation,kokubo2002formation}. Given this distribution, the outer disk has more massive planetary embryos and in smaller numbers than the inner disk. We used two different values of parameter $x$, 1.5 and 2.5, for each disk (internal and external). In the case of $x = 1.5$, we have less massive embryos at the beginning of the disk whereas, with $x = 2.5$, it results in a more massive disk in its initial part, see Figures \ref{fig:ini_form_z1} and \ref{fig:ini_form_z2}.  
We assume that embryos gravitationally interact with each other and with all other bodies of the system, in the case of planetesimals, they do not interact with each other but are allowed to interact with all other bodies. The eccentricity of all bodies was chosen randomly between $0-0.01$, the orbital inclination between 10$^{-4}-10^{-3}$ degrees and the mean anomaly between 0$^\circ$ and 360$^\circ$. The longitudes of the ascending nodes and the arguments of the periastron of all bodies were initialized with zero. 
For each of the disks, we have two distinct values of $x$ and in each of these cases, we performed 10 simulations with slightly different initial conditions for the protoplanetary embryos and planetesimals. Thus, there is a total of 20 simulations for each disk. Table \ref{tab:number_dist} shows the average number of protoplanetary embryos and planetesimals for the simulations of both disks with the two values of $x$. During the simulations, bodies that have a heliocentric distance of less than 0.1 $au$ or greater than 10 $au$ were removed from the system. All simulations were integrated by 100 Myr.

\begin{table}
\centering
\caption{Number of planetesimals and protoplanetary embryos distributed on the internal and external disks with the two values of the $x$ parameter.}
\label{tab:number_dist}
\begin{tabular}{cccccc}
\cline{2-6} \cline{2-6}
   & \multicolumn{2}{c}{Inner} &  & \multicolumn{2}{c}{Outer} \\ \cline{2-3} \cline{5-6} 
   & $x=1.5$         & $x=2.5$         &  & $x=1.5$         & $x=2.5$         \\ \hline
Embryos & 40          & 41          &  & 17          & 17          \\
Planetesimals & 480         & 480         &  & 480         & 480   \\
\hline
\hline
\end{tabular}
\end{table}

\subsection{Results and discussion}

\begin{figure*}
    \centering
	\includegraphics[scale=0.65]{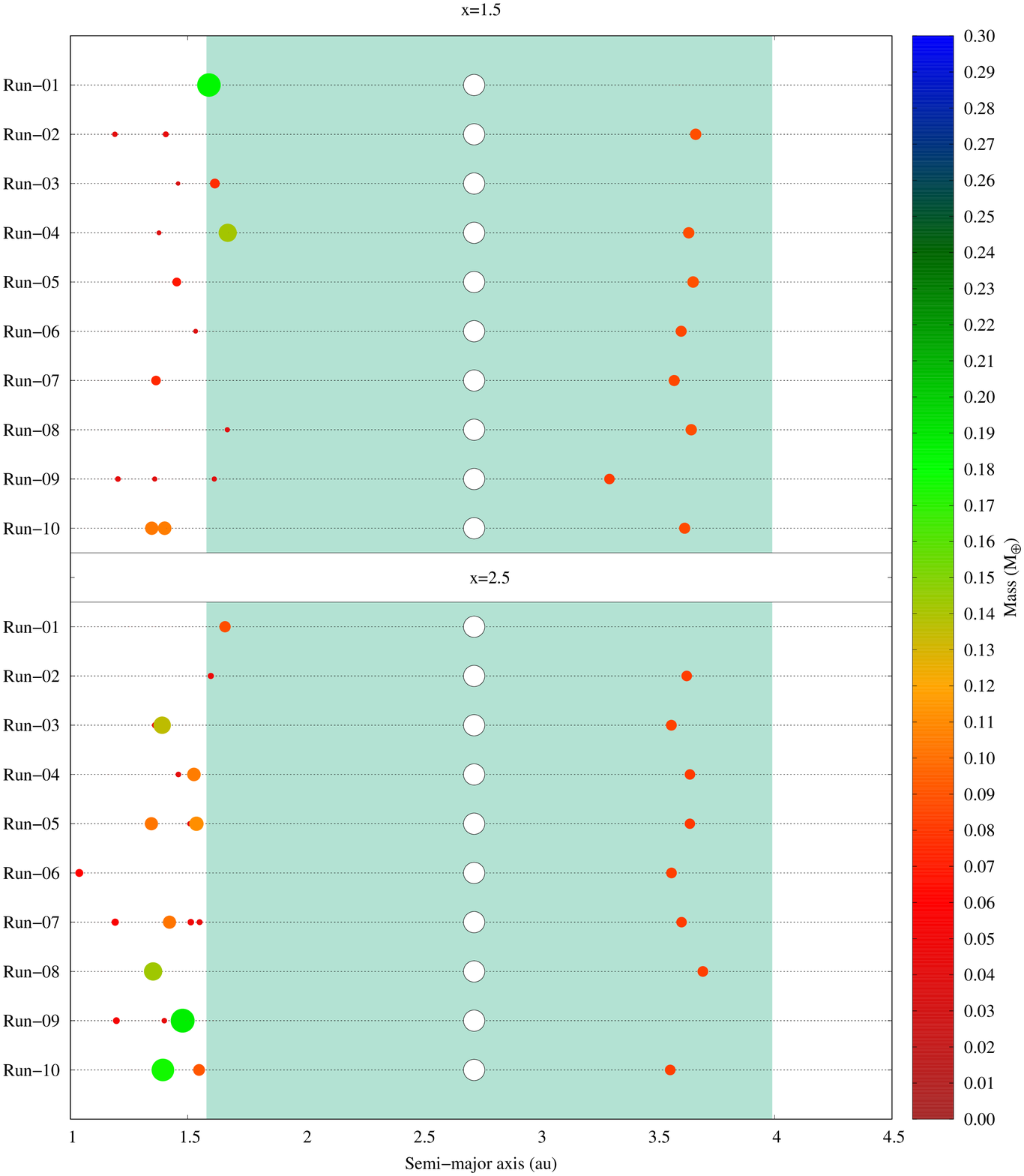}
    \caption{Final configuration of the 40 simulations performed. The figure shows the final mass of the surviving bodies in 100 Myr. In the center in white is the host planet \textit{Kepler-1647b} of the system in its current position. Each colored circle represents a body and its size is relative to its mass, except for the host planet. The green shaded region represents the system's HZ.}
    \label{fig:final_results}
\end{figure*}

Our numerical results of planetary formation are shown in Figure \ref{fig:final_results}. Two different disk mass distribution profiles were used in two regions considered stable within the HZ totaling 20 simulations, 10 with each profile in each of these regions. Thus, we will discuss the results of these simulations separately below.

\begin{figure*}
    \centering
	\includegraphics[scale=0.8]{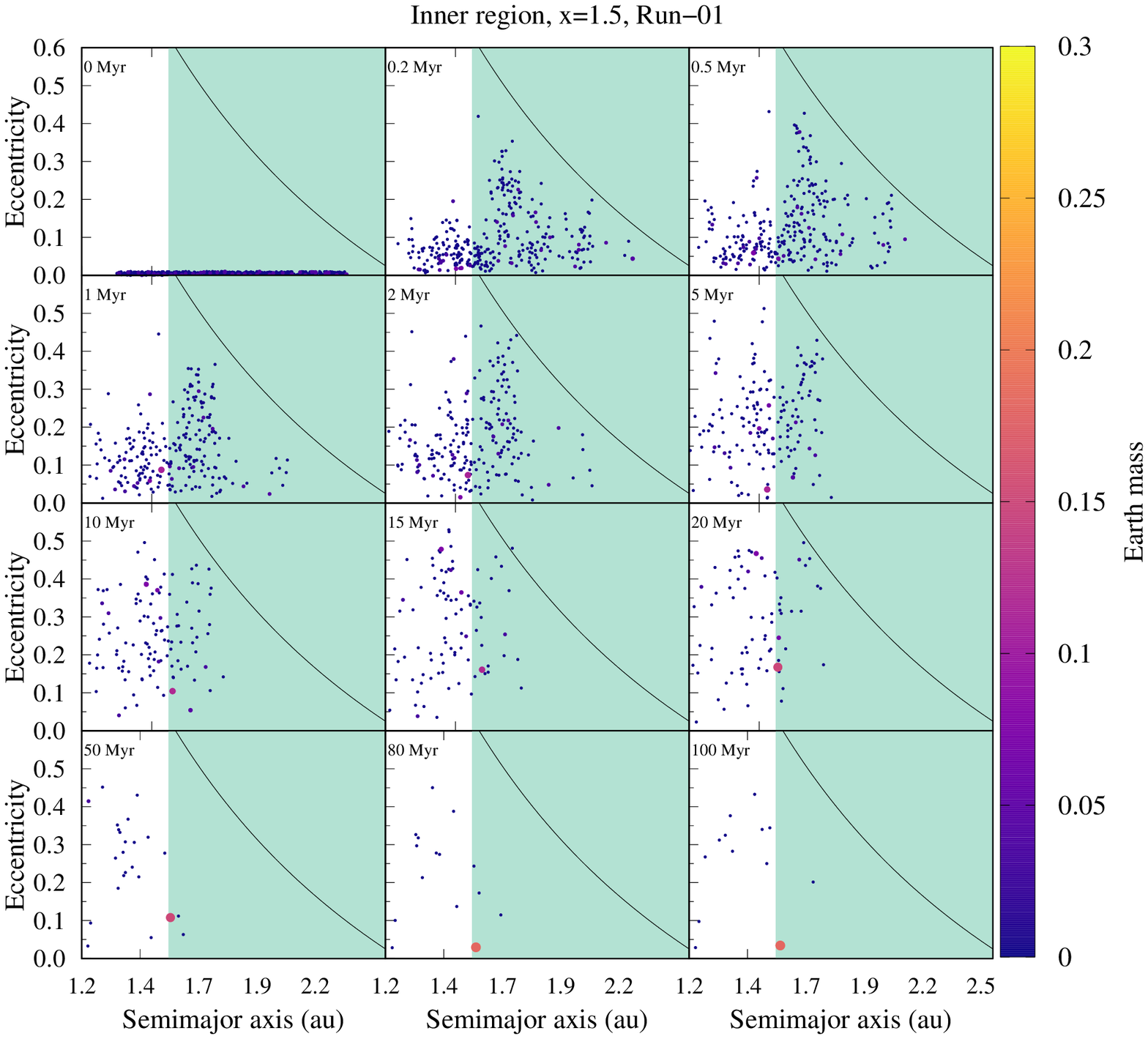}
    \caption{Snapshots in time of the dynamic evolution of protoplanetary embryos and planetesimals in the inner region with x=1.5 for the case of Run-01. The horizontal and vertical axes are the semi-major axis and the eccentricity, respectively. The coloured circles represent the embryos and planetesimals and their sizes are proportional to their masses. The blue lines represent the pericentre of the planet as a function of the apocentre of the particles, given by $a=[a_p(1-e_p)]/(1+e)$. The shaded region in green represents HZ of the system.}
    \label{fig:z1_15_1}
\end{figure*}

\begin{figure*}
    \centering
	\includegraphics[scale=0.8]{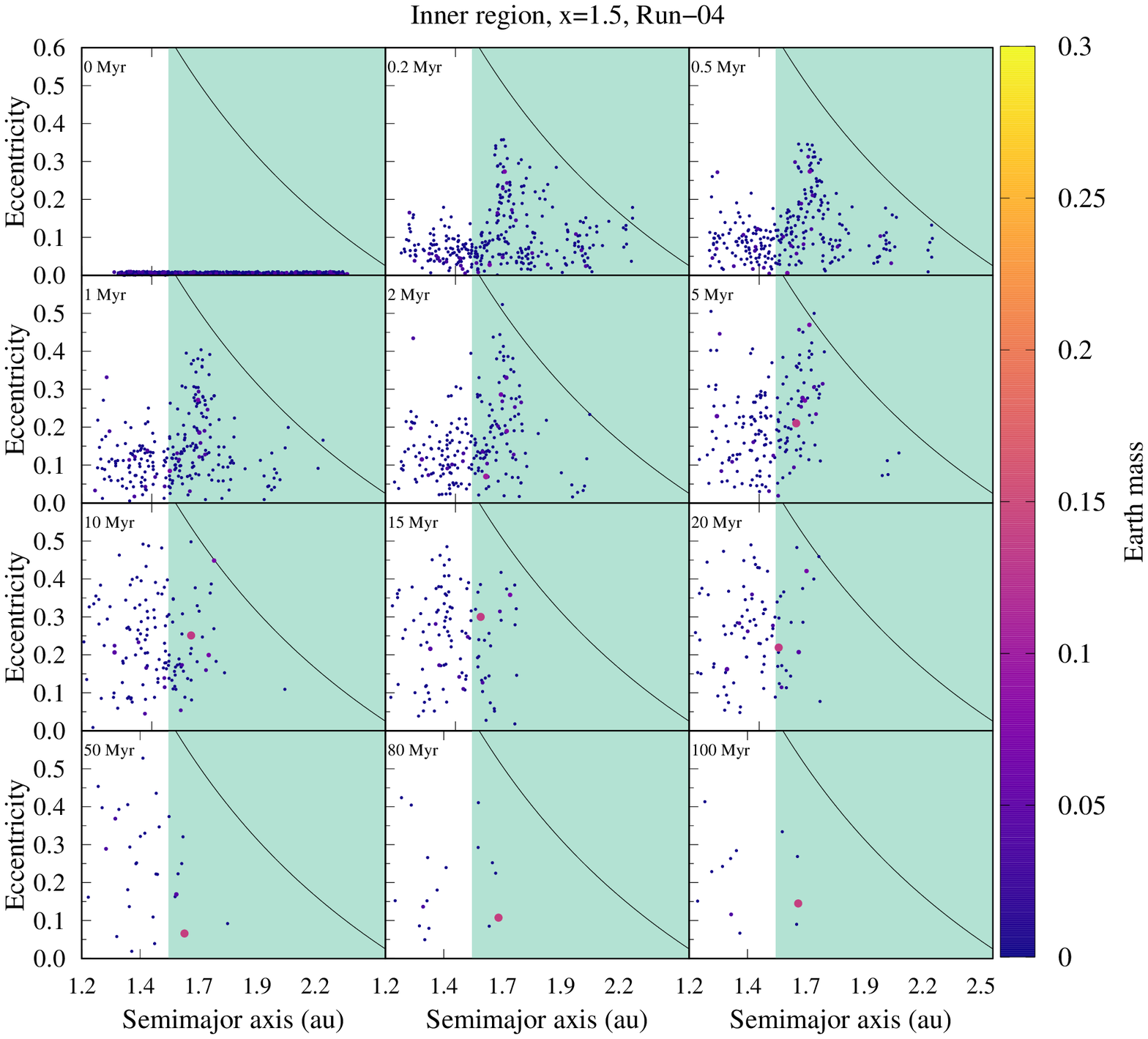}
    \caption{Snapshots in time of the dynamic evolution of protoplanetary embryos and planetesimals in the inner region with x=1.5 for the case of Run-04. The horizontal and vertical axes are the semi-major axis and the eccentricity, respectively. The coloured circles represent the embryos and planetesimals and their sizes are proportional to their masses. The blue lines represent the pericentre of the planet as a function of the apocentre of the particles, given by $a=[a_p(1-e_p)]/(1+e)$. The shaded region in green represents HZ of the system.}
    \label{fig:z1_15_4}
\end{figure*}

\begin{figure*}
    \centering
	\includegraphics[scale=0.8]{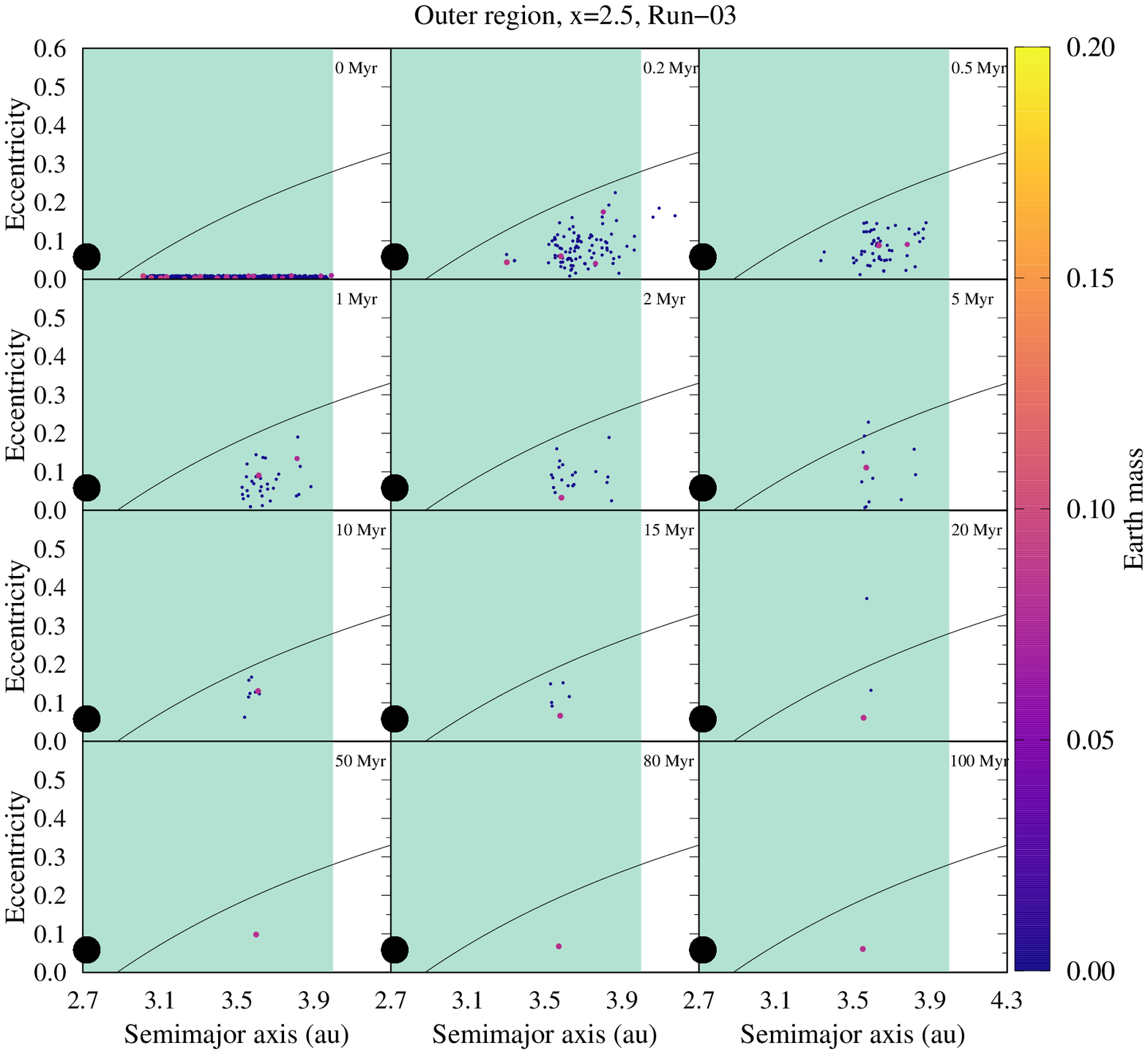}
    \caption{Snapshots in time of the dynamic evolution of protoplanetary embryos and planetesimals in the outer region with x=2.5 for the case of Run-03. The horizontal and vertical axes are the semi-major axis and the eccentricity, respectively. The coloured circles represent the embryos and planetesimals and their sizes are proportional to their masses, except the planet that is represented in black. The black line is the apocentre of the host planet of the system in function of the pericentre of the particle, given by $a=[a_p(1 + e_p)]/(1-e)$. The shaded region in green represents HZ of the system.}
    \label{fig:z2_25_3}
\end{figure*}
\subsubsection{Inner disk}

As previously mentioned, 20 numerical simulations were carried out with two different mass distribution profiles within this internal region. More precisely, it extends from 1.35 to 2.35 $au$ of the center of mass of the system.

In this case, protoplanetary embryos are less massive than in the outer region and consequently more numerous, as can be compared in Figures \ref{fig:ini_form_z1} and \ref{fig:ini_form_z2}. In the case where x = 1.5, the mass of the embryos is increasing along the disk, while in the case of x = 2.5, the mass is decreasing. This fact is made more important by the fact that the giant planet is present on the outer edge of the disk. Since this planet has approximately 1.5 masses of Jupiter, the disturbance caused by it on the disk is quite considerable. Figure \ref{fig:final_results}, shows that in all simulations of this region, the surviving bodies have a semi-major axis < 1.7 au. 

In addition to the proximity of a very massive planet, another factor that causes a lot of disruption in the disk are the internal resonances with the planet. Within this disk, four important internal resonances are located, as can be seen in Figure \ref{fig:zoon}. One being of the second-order (5:3) and other three of first-order (2:1, 3:2 and 4:3). These resonances increase the eccentricity of the bodies causing them to cross the planet's orbit. These intersections cause close encounters between the disc material and the planet, causing collisions with the planet or ejections of the bodies.

Even though the area of this disk is not large, which favors collisions taking into account the number of bodies on the disk, these resonances contribute to them being ejected before any massive body is formed. Figure \ref{fig:z1_15_4} shows that in the first hundreds of thousands of years, the eccentricity of the bodies varies widely, going from almost circular to approximately 0.4 in 0.2 Myr. Mainly around 1.71 au, where the 2:1 internal resonance is located.

Although the difference is small, the final mass of the disk is larger in the case where x = 2.5. This is due to the fact that with this profile, the most massive embryos are closer to the inner edge. In this way, they have a greater chance of surviving in the simulation. Overall, the final configurations are very similar in this region.

These results show that the effects caused by the giant \textit{Kepler-1647b} inside the disk make it impossible for a planet with the size of the Earth to be formed inside the HZ. However, this does not prevent a planet formed in another position in the system from being placed by migratory processes within the HZ. Our stability tests and planetary formation simulations show that specifically in this part of the HZ, its inner edge has stability for this to occur.

\subsubsection{Outer disk}

Likewise, we performed twenty simulations in the outer region, ten of them with x = 1.5 and another ten with x = 2.5. Despite having the same width, from 3 to 4 $au$, the area of this region is considerably larger, and the number of bodies within this disk is smaller.

In this case, bodies that have the smaller semi-major axis are closer to the host planet, and thus are more vulnerable to close encounters with it. The disc has three external resonances to the planet along its extension. Two of which are first-order (3:4 and 2:3) and one of second-order (3:5). These resonances contribute to the increase in the eccentricity of the disc bodies, as can be checked in Figure \ref{fig:zoon}. This increase in eccentricity causes the bodies to intersect with the apocentre of the planet, increasing the chances of close encounters that can cause both an increase in the semi-major axis and consequently ejections as well as collisions with it.

Even though the mass is the same in the two discs, the mass distribution of the protoplanetary embryos used makes their individual mass greater and, consequently, the amount of bodies is less. This decreases the probability that collisions will occur and consequently also decrease the chances of a massive body being formed. This is because the area is considerably larger than the one previously studied, which makes the time scale for collisions to occur to be greater. Looking in Figure \ref{fig:z2_25_3} we can note that most bodies are quickly ejected right at the beginning of the simulations, even before collisions occur.

Regardless of the value of parameter x, the result is quite similar, as we can check in Figure \ref{fig:final_results}. With x = 1.5, in two simulations (Run-01 and Run-0.3) none body survived. In all others, a protoplanetary embryo remained with its same initial mass. The same occurs with x = 2.5, in two simulations none body survives (Run-01 and Run-09) and in the others, only one body survived until the end of the integration. Except for the simulations that no body survived and the Run-09 case with x = 1.5, the surviving bodies are positioned between the external resonances 2:3 and 3:5 (see Figures \ref{fig:zoon} and \ref{fig:final_results}).

As in the previous disk, our results show that it is not possible to form an Earth-sized planet within this part of HZ. However, this does not prevent a planet formed in another region from being placed inside that region between the external resonances 2:3 and 3:5, in the giant's planet migratory processes.

\section{Stability of an Earth-type exomoon}
\label{subsec:exomoon}

As the planet \textit{Kepler-1647b} is within the habitable zone, in addition to the coorbital bodies, its satellites or exomoons will also be.
Some works have already studied the stability of an exomoon around a giant planet for long periods. In \cite{10.1111/j.1365-2966.2006.11104.x} through simulations of the three-body restricted elliptical problem, it is shown that a disk around a planet has an outer edge in 0.4895 Hill radius of the host planet of the system. However, these works are for systems that have only one star.

In the work of \cite{hamers2018stability}, the limits of stability of a moon around circumbinary planets discovered by the \textit{Kepler} probe are explored. It shows that the presence of a stellar companion affects these limits and that the stability boundary is well described by the location of the $1:1$ mean motion commensurability (MMC) with the stellar binary. However, it is also shown that in the case of the \textit{Kepler-1647} system, this effect is weak because the planet's semi-major axis is large. Thus, a simplification replacing the two stars with just one with the sum of their masses, does not produce major differences in results.

Therefore, through this simplification, we performed a stability test of a planet with a mass and size equal to the Earth around the planet Kepler-1647b. For this, we use the computational package of \textsc{MERCURY} n-bodies \citep{chambers1999hybrid} with the \textit{Bulirsch–Stoer} integrator.

\subsection{Numerical simulations: EXOMOON}

The parameters of the stars and the host planet used in the simulations can be checked in Table \ref{tab:binarydata}. In each simulation, in addition to the stars and the planet, there was also a satellite (exomoon) with mass and radius equal to that of Earth. We set up a grid of eccentricity per major axis for these satellites. 

The semi-major axis of the exomoons ($a_{exo}$) ranged from 1.5 radius of the planet ($r_p$) to 0.8 radius of Hill of the planet ($r_{H,p}$ $\approx$ 0.165 au) with an increment of 0.005 $r_{H,p}$. The eccentricity ($e_{exo}$) varied from 0.0 to 0.9 with an increase of 0.005, thus totaling 28,800 simulations. In all of them, the argument of the pericenter, ascending node and mean anomaly elements was set equals to zero. The final simulation time was $10^{4}$ orbital periods ($\approx$ 650 days) of an exomoon at 1 $r_{H,p}$. 

We consider as ejections in our simulations, bodies with a semi-major axis greater than 1 Hill radius of the planet and less than 1 radius of the planet, assuming the planet as a reference center.

\subsection{Results: EXOMOON}

\begin{figure}
    \centering
    \includegraphics[width=\linewidth]{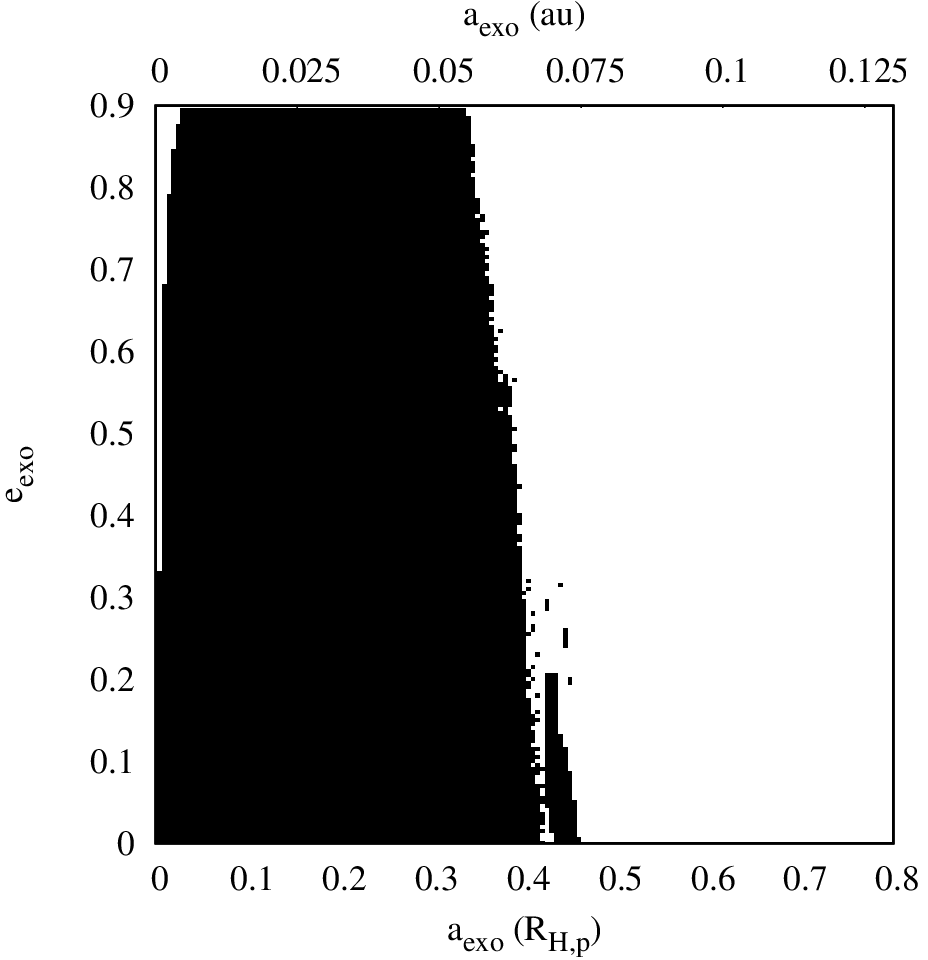}
    \caption{Simulation grid of the semimajor axis by the eccentricities of the Earth-type satellites around the planet \textit{Kepler-1647b}. In black, they are the exomoons that survived the total integration time, while in white they are the ones that were ejected.}
    \label{fig:exomoon}
\end{figure}

Figure \ref{fig:exomoon} shows in black the Earth type that survived after $10^4$ orbital periods. Blank are the particles that have been ejected from the system. In it, the planet Kepler-1647b is located at the origin. The stability limit here is approximately 0.4 Hill radii in the circular case, as can be seen in the Figure \ref{fig:exomoon}. In addition, we can also perceive an island formed beyond that limit. This result, even if by approximation, shows that the planet can house a planet the same size as the Earth inside the HZ as a satellite. The origin of this planet can be due to an in situ formation, or even a capture.

\section{Final Remarks}
\label{sec:conclusion}

In the present work, we used numerical simulations to investigate the possibility of a planet with a mass similar to the Earth`s to be formed within the HZ of the circumbinary system \textit{Kepler-1647}. Among all the binary systems with confirmed planets, this one has the most massive with \textit{P-type} orbit and the widest HZ. These characteristics make it one of the most complex systems of its kind and provide ample possibilities for exploring the formation of planets capable of harboring life.

For this purpose, we first computed the system's HZ limits. With the limits of HZ found, the second part of our work focused on studying stability within that zone. This test showed that the system has three stable sub-regions within HZ. One of them internal to the planet, another co-orbital to the host planet of the system and other external to the planet. This test shows us that there are several orbital resonances located along the particle disk. These resonances create disc gaps similar to what we see in the case of \textit{Kirkwood gaps} \citep{kirkwood1867meteoric,dermott1981dynamics}.

As the co-orbital region has many peculiarities concerning the other two regions, we explore this sub-region more fully. We used the initial conditions of the surviving particles of this specific test to perform simulations of planetary formation. Using eight values of total disk mass we ran 80 different cases, ten for each value of total mass randomly varying the mass of the bodies by 200 Myr. The results of these simulations showed that it is impossible to form a body with a mass close to that of the Earth co-orbitally to the system planet. However, our results also show that the system has conditions to have bodies co-orbital to the planet similar to the Trojans asteroids in our Solar system. 

For the other two regions, we use two distinct surface density profiles to explore planetary formation. For each distinct value of the density profile, we performed 10 simulations, totaling 40 adding the two regions. In the case that we have a disk density profile equal to 1.5, the largest planet formed was in the inner region (Run-01). This planet has approximately 0.2 masses of Earth (1.8 masses of Mars). In the event that this value is equal to 2.5, the largest planet formed is outside the HZ (0.25 Earth's mass).

As the planet is located in the center of the HZ, a hypothetical satellite from that planet would consequently be inside the HZ. Since we have shown that a planet the size of Earth cannot form within the system's HZ, we looked for moons. Thus, we performed several computer simulations to find the possible locations where a satellite of the Earth size can be stable. As a result, we found that the planet is capable of harboring an Earth analogue up to approximately 0.4 Hill radius  from the center of the planet.

Note added in proof: In the work \cite{georgakarakps2021} is applied an analytical approach concerning the HZ for the \textit{Kepler-1647} system, where they conclude that it is unlikely to host habitable worlds. 

\section*{Acknowledgements}

The work was carried out with the support of the \textit{Improvement Coordination Higher Education Personnel} - Brazil (CAPES) - Financing Code 001, with the support of the \textit{São Paulo Research Foundation} (FAPESP - proc. 2015/50122-0 and  proc. 2016/24561-0) and to \textit{National Council for Scientific and Technological Development} (CNPq) via grant 305210/2018-1. The research had computational resources provided by \textit{Center for Mathematical Sciences Applied to Industry (CeMEAI)}, funded by FAPESP (proc. 2013/07375-0).

\section*{Data Availability Statements}
The data underlying this article will be shared on reasonable request to the corresponding author.

%%%%%%%%%%%%%%%%%%%%%%%%%%%%%%%%%%%%%%%%%%%%%%%%%%
\section*{ORCID iDs}
G. O. Barbosa \orcidicon{0000-0002-1147-2519} \href{https://orcid.org/0000-0002-1147-2519}{https://orcid.org/0000-0002-1147-2519}\\
O. C. Winter \orcidicon{0000-0002-4901-3289} \href{https://orcid.org/0000-0002-4901-3289}{https://orcid.org/0000-0002-4901-3289}\\
A. Amarante \orcidicon{0000-0002-9448-141X} \href{https://orcid.org/0000-0002-9448-141X}{https://orcid.org/0000-0002-9448-141X}\\
E. E. N. Macau \orcidicon{0000-0002-6337-8081} \href{https://orcid.org/0000-0002-6337-8081}{https://orcid.org/0000-0002-6337-8081}
\bibliographystyle{mnras}
\bibliography{references} % if your bibtex file is called example.bib
%\begin{thebibliography}{99}
%\end{thebibliography}
%%%%%%%%%%%%%%%%%%%%%%%%%%%%%%%%%%%%%%%%%%%%%%%%%%
% Don't change these lines
\bsp	% typesetting comment
\label{lastpage}
\end{document}